\documentclass[aps,prl,groupedaddress]{revtex4-2}
\usepackage[dvipsnames]{xcolor}
 \usepackage{graphicx}
 \usepackage{amsmath}
\usepackage{amssymb}
\usepackage{cancel}
\usepackage[normalem]{ulem}
\usepackage{amsmath}
\newcommand{\stkout}[1]{\ifmmode\text{\sout{\ensuremath{#1}}}
\else\sout{#1}\fi}

\usepackage{mathptmx}
\newcommand{\car}{\eta_{\rm C}^{}}
                                          
\begin{document}
%\pagecolor{Cyan}
\title{Coupled autonomous thermal machines and    
efficiency at maximum power}
 \author{Ramandeep S. Johal} 
 \email[e-mail: ]{rsjohal@iisermohali.ac.in}
 \affiliation{ Department of Physical Sciences, 
 Indian Institute of Science Education and Research Mohali,  
 Sector 81, S.A.S. Nagar, Manauli PO 140306, Punjab, India} 
 \author{Renuka Rai}
	\email[e-mail: ]{rren2010@gmail.com}
	\affiliation{Department of Applied Sciences, 
	University Institute of Engineering and Technology (U.I.E.T), 
	Panjab University, Sector 25, Chandigarh 160014, India}
 \date{\today}
\begin{abstract}

We show that coupled autonomous thermal machines, in the presence
of three heat reservoirs and following a global linear-irreversible
description, provide a unified framework to accomodate the variety
of expressions for the efficiency at maximum power (EMP). The
efficiency is expressible in terms of the Carnot efficiency
of the global set up if the intermediate reservoir temperature
is an algebraic mean of the hot and cold temperatures. 
We give an explanation of the universal properties of
EMP near equilibrium in terms of the properties 
of symmetric algebraic means. For the case of broken
time reversal symmetry, a universal second order
coefficient of 6/49 is predicted in the series expansion of EMP, 
analogous to the 1/8 coefficient in the time-reversal symmetric case.
\end{abstract}
\newpage
\maketitle
{\it Introduction}:
We observe that engines in the real world involve fluxes of matter and energy
and undergo processes with finite rates. Linear-irreversible 
thermodynamics is by far the simplest phenomenological theory 
that assumes the fluxes  
 to be proportional to the small thermodynamic forces driving them 
\cite{Onsager1931a}. 
Heat engines based on this premise and other auxilliary assumptions bound  the 
efficiency at maximum power (EMP), e.g. as  $\eta_{\rm C}^{}/2$  
\cite{Broeck2005}, 
where $\eta_{\rm C}^{}$ is the Carnot efficiency.
Other irreversible models \cite{Curzon1975, Chen1989, 
Schmiedl2008, Esposito2010prl, Moreau2012, Benenti2011, Seifert2013,  
 Izumida2012, Broeck2013, Johal2018}   may 
predict  EMP that goes beyond the linear 
response result .
These expressions for EMP  are usually model-specific (see Table 1 for a few 
examples),  
although they fall within certain bounds, as for example: 
\begin{equation}
\frac{\eta_{\rm C}}{2} \leq \eta^{}_{\rm MP} \leq \frac{\eta_{\rm C}}
{2 - \eta_{\rm C}}.
\label{bnd_sc}
\end{equation}
Invariably, expressions of $\eta^{}_{\rm MP}$  exhibit a dependence
 on the ratio of cold to hot reservoir temperatures ($T_c/T_h$), 
an important feature also of the Carnot efficiency, $\eta_{\rm C}^{} = 1 - 
T_c/T_h$. 
Other universal or model-independent features can be identified at small values
of $\eta_C$ (near-equilibrium situations), whereby the EMP
satisfies the series expansion: $\eta^{}_{\rm MP}  \approx \eta_{\rm C}^{}/2 + 
\eta_{\rm C}^{2}/8 
+ {\cal O}\left[ \eta_{\rm C}^{3}  \right]$.
Here, the first order coefficient (1/2) corresponds to the linear-response 
behavior, while the second-order coefficient (1/8) has been analyzed in 
terms of a certain left-right symmetry of the specific model  
\cite{Lindenberg2009, Lindenberg2010, Johal2018,Johal2019}.
The fact that many proposed models do 
show universal features in EMP, suggests  
the possibility of a generic thermodynamic model 
that might incorporate the various expressions
within a single framework \cite{comment_Johal2018}.
However,  to the best of our knowledge, 
there exists no scheme for autonomous machines 
that may accomodate the myriad expressions
for EMP in a unified framework while accounting for its universal
features. 
\begin{figure}
 \includegraphics[width=8cm]{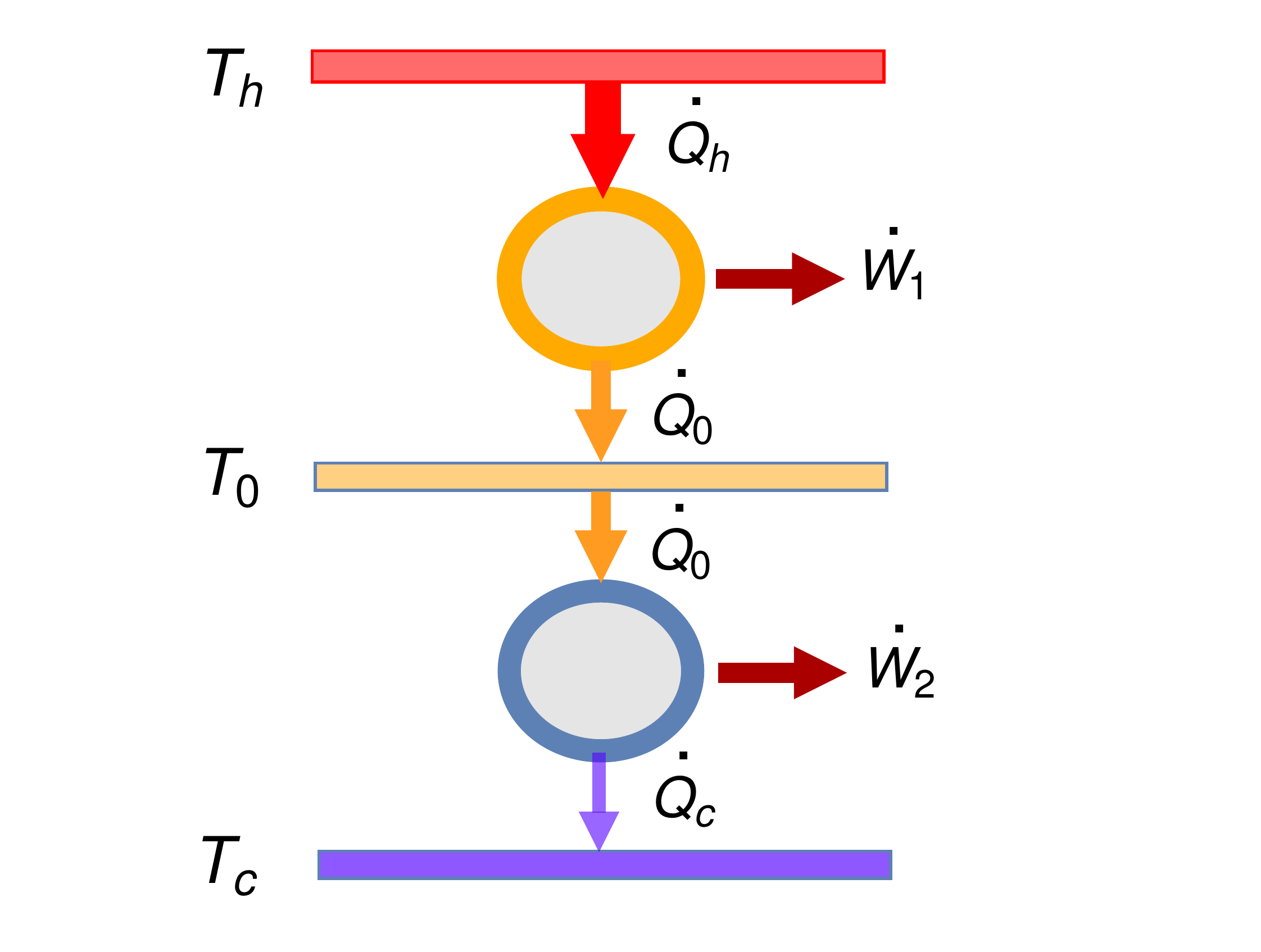}
 \caption{Two autonomous heat engines tightly coupled via 
 a third heat reservoir at temperature $T_0$,
 which satisfies $T_c \le T_0  \le T_h$.
 The total power output is $\dot{W} = \dot{W}_1  + \dot{W}_2$.
 }
\end{figure}
\par\noindent
In this paper, we analyse the global performance of two autonomous
heat engines which are tightly coupled via a third
heat reservoir having a temperature intermediate between the hot and cold 
reservoirs (see Fig. 1).
Within a linear-irreversible framework, we optimize
the total power output and show that EMP is bounded
as $\eta^{*} \leq {\eta_{\rm C}^{}} \left({1+{T_c}/{T_0}}\right)^{-1}$,
where the upper bound is achieved under strong coupling (SC)  
condition.
The previous bound of $\eta_{\rm C}^{}/2$ is recovered for $T_0 = T_c$, but
can be breached for $T_0 > T_c$ .
Further, the requirement that  EMP depends only
on the ratio $T_c/T_h$, or equivalently upon $\eta_{\rm C}^{}$ 
\cite{comment_car},
requires that $T_0$ be expressed as a mean value of the 
hot and cold temperatures. Interestingly, specific choices of 
some common means  
for $T_0$ rather lead to well-known expressions
for the EMPs of two-reservoir heat engines  
(Table 1). This also attributes the above mentioned   
universal features to  EMP, if the choice is restricted to 
the so-called symmetric means. 
We also derive EMP for sub-optimal coupling and 
suggest a new universality class for EMP in the case of broken time 
reversal symmetry (TRS). More precisely, in place of the universal 1/8 
coefficient in the series expansion of EMP, we derive a universal coefficient
of 6/49 for the case of broken TRS.
Finally, apart from the engine, we are able to optimize  the cooling power in 
the refrigerator mode---a goal which proves to be elusive in some of the 
previously studied models.
\begin{table}[ht]
\begin{ruledtabular}
\begin{tabular}{c c c c}
Mean & $T_0 \equiv {\cal M}(T_h,T_c)$ & $\eta_{u}^{*} = \car \left( {1+ 
\frac{T_c}{T_0}} \right)^{-1}$ & Physical model \\ \hline
& & & \\
Geometric & $\sqrt{T_h T_c}$ & $1-\sqrt{1-\car }$ & Ref. \cite{Curzon1975} \\
&&& \\   
 Harmonic \quad & $\dfrac{2 T_h T_c}{T_h + T_c} $\quad &  
 $\dfrac{2\car }{4- \car}$ & Refs. \cite{Chen1989, Schmiedl2008,
 Broeck2013, JohalRai2016}  \\
  & & & \\
% %
% %
Arithmetic \quad &  $\dfrac{T_h + T_c}{2}$ &  
$\dfrac{(2- \car)\car}{4 - 3\car}$ & Ref.  \cite{VarinderJohal2018}  \\
   & & & \\ 
 Logarithmic & $\dfrac{T_h - T_c}{\ln T_h -\ln T_c}$ & 
 $\dfrac{\eta_{\rm C}^{2}}
 {\car - (1-\car)\ln (1-\car)}  $ & Refs. \cite{Tu2008, Broeck2012, 
 WangTLS2013, Erdman2017} \\
 &&& \\
% %
 Lehmer & $\dfrac{T_{h}^{\sigma} +T_{c}^{\sigma} }{T_{h}^{\sigma-1}+ 
 T_{c}^{\sigma-1}}$ & 
$\dfrac{\car }{2-\dfrac{\car }{1+ (1-\car)^\sigma}}$ & $\sigma \in R$,  Ref. 
 \cite{Cavina2017} 
% %
 \end{tabular}
 \end{ruledtabular}
 \caption{The intermediate temperature ${T_0}$ as some well-known symmetric 
 means of $T_h$ and $T_c$, and the corresponding upper bound for EMP, 
$\eta^{*}_{u}$, obtained under strong coupling, where $\car = 1-T_c/T_h$. 
Various finite-time models derive these forms of EMP in the listed 
references.} 
 \end{table}

Based on Fig. 1, let us now consider the performance of 
the sub-engines. The reservoirs $T_h$ and $T_0$ are 
coupled via an autonomous engine 
leading to power output $\dot{W}_1 = \dot{Q}_{h} - \dot{Q}_{0}$, 
and a rate of entropy generation,  
$\dot{S}_1 = -\dot{Q}_{h}/T_h + \dot{Q}_{0}/T_0$, 
which can be written as:
\begin{equation}
 \dot{S}_1 = -\frac{\dot{W}_1}{T_0}+ {\dot{Q}_{h}}
\left(\frac{1}{T_0}-\frac{1}{T_h}\right).
\label{sdot1} 
\end{equation}
Similarly, reservoirs $T_0$ and $T_c$
are coupled via another such engine that leads
to power output $\dot{W}_2 = \dot{Q}_{0} - \dot{Q}_{c}$,
and a rate of entropy generation, 
$\dot{S}_2 = -\dot{Q}_{0}/T_0 + \dot{Q}_{c}/T_c$,
which can be written as:
\begin{equation}
\dot{S}_2 
= -\frac{\dot{W}_2}{T_0}+ {\dot{Q}_{c}}
\left(\frac{1}{T_c}-\frac{1}{T_0}\right).
\label{sdot2} 
\end{equation}
Since the two sub-engines  are  tightly coupled with each other,
the net heat flux exchanged with the intermediate 
reservoir is zero. 
Then,  $\dot{W}_1 + \dot{W}_2 = \dot{Q}_{h} - \dot{Q}_{c} = \dot{W}$, and  
$\dot{S}_1 + \dot{S}_2 = 
\dot{S}$ is written as:
\begin{equation}
\dot{S} = -\frac{\dot{W}}{T_0}+\dot{Q}_h \left( 
\frac{1}{T_0}-\frac{1}{T_h}\right)+\dot{Q}_c 
\left(\frac{1}{T_c}-\frac{1}{T_0} \right).
\label{sdt0}
\end{equation}
\par\noindent
Let us define:
$X_{h} = 1/T_0 - 1/T_h \geq 0$ and $X_{c} = 1/T_c - 1/T_0 \geq 0$,
so that ${X_h +X_c} = 1/T_c - 1/T_h$. 
Then, we can write Eq. (\ref{sdt0}) as
 \begin{align}
\dot{S} & = -\frac{\dot{W}}{T_0} + \cfrac{\dot{Q}_h X_h +
\dot{Q}_c X_c}{X_h +X_c}(X_h + X_c), 
\nonumber \\
& =  -\frac{\dot{W}}{T_0}+ {\dot{Q}_{\rm av}}
\left(\frac{1}{T_c}-\frac{1}{T_h}\right),
\label{sdotf}
\end{align}
where the average or effective thermal flux is given by
\begin{equation}
\dot{Q}_{\rm av} =(1-\omega)\dot{Q}_h + \omega \dot{Q}_c, 
\label{qav}
\end{equation}
with $\omega = X_c/(X_{h} + X_{c})$ satisfying $0\le \omega \le 1$.
In standard approaches, the reference reservoir is usually
chosen to be the coldest reservoir available, and so $T_0 = T_c$.
Within the present framework, the reference 
reservoir is an additional resource at $T_0$ and
the relevant thermal flux is the average value $\dot{Q}_{\rm av}$.
Finally, the total power flux is given as: $\dot{W} = F \dot{x}$,
where $F$ is the load and $\dot{x} \equiv \dot{x}_1 + \dot{x}_2$ is the 
total rate of displacement generated.

Now, assuming a linear-irreversible description at the level 
of global performance of the coupled engines, 
we identify the following 
flux-force pairs:
\begin{align}
{J}_1 &= \dot{x}, \qquad X_1 = - \frac{F}{T_0},
\label{j3mac}\\
{J}_2 &= \dot{Q}_{\rm av}, \quad X_2 = \frac{1}{T_c} - \frac{1}{T_h},
\label{j4mac}
\end{align}
so that the rate of entropy production is cast in a bilinear form 
$\dot{ S} =\sum_{i=1}^{2} {J}_i X_i$.
Secondly, the linear regime implies the 
flux-force relations of the form: $J_i = \sum_{j=1}^{2} L_{ij} X_j$, where 
$i=1,2$.
Here, the phenomenological coefficients $L_{ij}$ are assumed  fixed
due to the  small magnitudes of the forces.
Then, the second-law inequality 
imposes the following conditions:
\begin{equation}
L_{11}, L_{22} \geq 0, \qquad
4 L_{11} L_{22} \geq { \left( L_{12} + L_{21} \right)^2}.
\label{Lcond}
\end{equation}
We first assume the principle of microscopic 
time-reversal symmetry (TRS) which allows the use of 
Onsager reciprocity relation:
$L_{21} = L_{12}$.
In this case, the third inequality above 
reduces to $ L_{11} L_{22} \geq  L_{12}^{2}$.
This makes it convenient to define a measure, 
$q = L_{12}/\sqrt{L_{11} L_{22}}$, 
for the coupling strength between thermodynamic forces,
which satisfies:   $-1 \leq q \leq +1$.

So, the constitutive relations for the fluxes in Eqs. (\ref{j3mac})
and (\ref{j4mac}) can be written in the following form. 
\begin{align}
\dot{x}  &= -{L}_{11} \frac{F}{T_0} + {L}_{12} X_2,
\label{j1glo} \\
\dot{Q}_{\rm av} &= -{L}_{12}\frac{F}{T_0}  + {L}_{22} X_2.
\label{j2glo}
\end{align}
From  (\ref{qav}) and (\ref{j2glo}),
we can derive the following relations:
\begin{align}
\dot{Q}_{h} &= -{L}_{12}\frac{F}{T_0}  + {L}_{22} X_2
+ \omega \dot{W},
\label{qh} \\
\dot{Q}_{c} &= -{L}_{12}\frac{F}{T_0}  + {L}_{22} X_2
-(1-\omega) \dot{W}.
\label{qc}
\end{align}
\par\noindent
{\it Optimization of power output}:
By using Eq. (\ref{j1glo}), we optimize the power output, 
$\dot{W} = F\dot{x}$, with respect to the load $F$.
The optimal load is obtained at   
$F^* = L_{12} T_0 X_2/2 L_{11}$.
The optimal power, $\dot{W}^* \equiv \dot{W} (F^*)$, 
is given by:
\begin{equation}
\dot{W}^* = \frac{L_{12}^{2} T_{0}^{} X_{2}^{2}}{4 L_{11}}.
\label{mp}
\end{equation}
Similarly, the hot flux, $\dot{Q}^{*}_{h} \equiv \dot{Q}_h (F^*)$, 
is obtained from Eq. (\ref{qh}) as:
\begin{equation}
\dot{Q}_{h}^{*} = \left[ 1+ \frac{q^2}{4}\left(\frac{T_0}{T_c} -3\right) 
\right] L_{22} X_2.
\label{qhstar}
\end{equation}
Then, the efficiency at maximum power (EMP), 
$\eta^{*}_{} =  \dot{W}^*/{\dot{Q}_{h}}^{*}$,
is evaluated to be:
\begin{equation}
\eta^{*}_{} =  {\eta_{\rm C}}
\left[ {1+ \left(\frac{4-2q^2}{q^2} -1 \right)  \frac{T_c}{T_0} }
\right]^{-1}.
\label{etas}
\end{equation}
For given reservoir temperatures, the EMP can be varied
by tuning the coupling strength 
$q$, but it remains bounded as:
\begin{equation}
  0 \leq \eta^{*}_{} \leq {\eta_{\rm C}^{}} 
\left ({1+\frac{T_c}{T_0}}\right )^{-1}  \equiv 
\eta_{u}^{*}
\label{etau}
\end{equation}
 where the upper bound is saturated 
for strong coupling ($q^2=1$).

Next, we address universal properties of EMP in the
context of our coupled model. 
The previous studies on universality of EMP were 
mostly carried out on two-reservoir set ups, where 
EMP is  obtained as a function of $T_c/T_h$. In the present model, 
with three reservoirs, the EMP depends on two ratios
involving three temperatures, as in Eq. (\ref{etas}). 
Now,  $T_0$ may be assigned some numerical value
in the  interval $[T_c,T_h]$. However, as we show
in the following, when $T_0$ is expressed as
an algebraic mean of $T_h$ and $T_c$, 
then the EMP depends only on $T_c/T_h$
and we can establish a comparison with the EMP
of two-reservoir models.

Let  ${\cal M}(a,b)$ define an algebraic mean of two real numbers $a, b >0$, 
which satisfies $\text{min}[a,b] < {\cal M}(a,b) < \text{max}[a,b]$.
So, we define ${\cal M}(a,a) = a$.
Further, ${\cal M}$ is 
a homogeneous function of its arguments, satisfying
 ${\cal M}(\lambda a, \lambda b) = \lambda {\cal M}(a,b)$, for 
 all real $\lambda$. Thus, we can write 
 ${\cal M}(a,b) = a {\cal M}(1,b/a)$. 
Assuming $T_0$ to be such a mean 
of  hot and cold temperatures, i.e. $T_0 \equiv  {\cal M}(T_h, T_c)$, 
we can write: 
$T_0 \equiv T_h {\cal M}(1, T_c/T_h) = T_h {\cal M}(1,1-\eta_{\rm C})$. 
In other words, $\eta^*$ of Eq. (\ref{etas} ) becomes 
a function only of $\eta_{\rm C}$, or the ratio of cold to hot temperatures.

Then,
for a small difference between the hot and cold temperatures
($\eta_{\rm C}$ as small parameter), we may develop ${\cal M}$ as
a Taylor series in $(-\eta_{\rm C})$:
\begin{equation}
{\cal M}(1,1-\eta_{\rm C}) = 1 + a_1 (-\eta_{\rm C}) + 
a_2 (-\eta_{\rm C})^{2} + {\cal O}[\eta_{\rm C}^{3}],
\label{mec}
\end{equation}
where the coefficients $a_1, a_2,...$ are
determined by the form of the given mean \cite{comment_taylor}.
The corresponding series expansion of Eq. (\ref{etas}) is then given by:
\begin{equation}
\eta^{*}_{} =  \frac{q^2}{4 - 2 q^2} \eta_{\rm C}
+ (1-a_1) \frac{(4 - 3 q^2)q^2}{(4 - 2 q^2)^2}\eta_{\rm C}^{2}
+ {\cal O} [\eta_{\rm C}^{3} ].
\label{etsq}
\end{equation}
The first order term above is the same as 
for a two-reservoirs (hot and cold) set up \cite{Broeck2005}, being 
independent of the
intermediate temperature $T_0$. For $q^2=1$, 
this term yields the half-Carnot value.
The coefficient of the second-order term   
depends on $q^2$ as well as on $a_1$ which is a characteristic
of the mean $T_0$ (see Eq. (\ref{mec})). Remarkably, if $T_0$ is  a
{\it symmetric} mean, i.e. having the property 
 ${\cal M}(T_h, T_c) = {\cal M}(T_c, T_h)$, then $a_1 =1/2$ \cite{comment_a1}, 
 and  we may rewrite Eq. (\ref{etsq}) as:
\begin{equation}
\eta^{*}_{} = \beta \eta_{\rm C} + \frac{\beta(1-\beta)}{2} \eta_{\rm C}^{2}
+ {\cal O} [\eta_{\rm C}^{3}],
\label{etab}
\end{equation}
where  $\beta \equiv {q^2}/{(4 - 2 q^2)}$ and  $0 \leq \beta \leq 1/2$. 
Thus, we have a universal relation between the first and second order
coefficients, which is valid for any choice of the symmetric mean $T_0$. 
In particular, for models with SC, $\beta = 1/2$,  and thus  
 we obtain 1/8 as the  second-order coefficient, 
analogous to the two-reservoirs case \cite{Lindenberg2009}.

Interestingly, 
many known expressions for EMP 
can be derived by assigning a specific mean to $T_0$.
The few examples of Table I pertain to the 
scenario $q^2=1$, for which Eq.   (\ref{etas})
yields  $\eta^{*}_{} = \car \left({1+ {T_c}/{T_0}} \right)^{-1}$. 
Upon comparison between this formula and a known expression for  
EMP, the corresponding $T_0$ may be inferred.

As another example, a tandem construction of linear-irreversible
engines \cite{Broeck2005} leads to the EMP, $\eta^{*}_{} = 1 
-(1-\eta_C)^{\beta}$. 
Comparing this expression for EMP
with Eq. (\ref{etas}), we obtain:
\begin{equation}
 T_0 = \cfrac{\beta -1}{\beta} \;  \cfrac{ T_{h}^{\beta} -  
T_{c}^{\beta}} 
 {T_{h}^{\beta-1} -  T_{c}^{\beta-1}   },
\end{equation}
a special case of the generalized mean  \cite{Stolarsky1975,Leach1978}. 
Due to $0 \leq \beta \leq 1/2$, $T_0$ is bounded as:  
$T_h T_c/T_{\rm  L} \leq T_0 \leq \sqrt{T_h T_c}$, with 
$T_{\rm L} = (T_h - T_c) /\log (T_h/T_c)$ as the logarithmic mean.
Here, CA-efficiency \cite{Curzon1975}
is obtained with $\beta=1/2$, for which $T_0 = \sqrt{T_h T_c}$. 

Further, it is not hard to find 
examples of asymmetric means, ${\cal M}(T_h, T_c) \neq {\cal M}(T_c, T_h)$, 
that can parametrize more general expressions of EMP.
Thus, the use of weighted harmonic mean: 
$T_0 = {T_h T_c}/[(1-\alpha) T_h + \alpha T_c]$ in Eq. (\ref{etas})
yields $\eta^{*}_{} = {\car }/({2- \alpha \car})$, where $0\leq \alpha \leq 1$.
The symmetric case of $\alpha =1/2$
has been already mentioned in Table 1.
The above expression has been derived in various models,  
where, for instance, the parameter
$\alpha$ may quantify the ratio of heat transfer coefficients \cite{Chen1989}
or dissipation constants  \cite{Broeck2013, Johal2019}
on the hot and cold sides of the engine. 
  
The basic framework can be easily generalized 
to scenarios with a broken TRS, for which the reciprocity
relation is no longer true, i.e. $L_{21} \neq 
 L_{12}$. Then, the second flux-force relation, Eq. (\ref{j2glo}), 
 reads as: 
$\dot{Q}_{\rm av} = -{L}_{21}{F}/{T_0}  + {L}_{22} X_2$.
Following an analogous derivation as for the time-symmetric case, the EMP 
is given as:
\begin{equation}
\eta^{*}_{{\cancel{\scriptscriptstyle{\text{ TRS}}}}} =  {\eta_{\rm C}}
\left[ {1+ \left( \frac{1-{\gamma}}{{\gamma}}    \right)  \frac{T_c}{T_0} }
\right]^{-1}.
\label{etasbr}
\end{equation}
Here, $\gamma \equiv x y/(4+2 y)$,
with  $x = L_{12}/L_{21}$ and $y = L_{12} L_{21}/ 
(L_{11}L_{22} - L_{12}L_{21})$ \cite{Benenti2011}. 
For $x=1$, 
we can write $y = q^2/(1-q^2)$ or $\gamma = \beta$,
and so Eq. (\ref{etasbr})
reduces to Eq. (\ref{etas}), thus recovering 
the results of the model satisfying TRS.
Secondly, note that for $T_0 = T_c$, results
of the previous studies \cite{Benenti2011,Seifert2013} are recovered, by which
$\eta^{*}_{{\cancel{\scriptscriptstyle{\text{ TRS}}}}} = \gamma \eta_{\rm 
C}^{}$. 
Thus, the presence of an additional reservoir at $T_0 > T_c$,
raises the EMP beyond  $\gamma \eta_{\rm C}^{}$. 

As noted in Refs.  
\cite{Benenti2011, Seifert2013},
for a given value of $x$, the parameter 
$\gamma$ lies in the range:
$ 0\leq \gamma \leq x^2/(4x^2-6x+4) \equiv \hat{\gamma}$. 
Since the EMP of Eq. (\ref{etasbr})
is a monotonic increasing function of
$\gamma$, so the optimal EMP is given by:
\begin{equation} 
\eta^{*}_{{\cancel{\scriptscriptstyle{\text{ TRS}}}}} = 
\eta_{\rm C}^{} \left[1+ 
\frac{1-\hat{\gamma}}{\hat{\gamma}} 
\cfrac{T_c}{T_0} \right]^{-1}.
\label{etasbrx}
\end{equation}
Clearly, for $x=1$, we obtain $\hat{\gamma} =1/2$,
recovering the results of   the 
strong-coupling
case, $\eta^* = \eta_{\rm C}^{}/ (1+T_c/T_0)$. 
Fig. 2 plots Eq. (\ref{etasbrx}) for different special cases.
As argued in Ref. \cite{Seifert2013}, the 
upper bound of EMP  can be breached in the case of 
broken TRS, yielding the optimal EMP  as
 $ 4 \eta_{\rm C}^{}/7 (> \eta_{\rm C}^{}/2)$. 
It is apparent from Fig. 2 that the intermediate
temperature $T_0$ helps to go beyond this result too and so the 
bound $4 \eta_{\rm C}^{}/7$ is rendered just as  
the {\it lower} bound. 

Although the exact expression for EMP depends on the 
specific form of $T_0$, we can inquire into the universal 
features just as for the case with TRS. 
For an arbitrary symmetric mean $T_0$, and in proximity to equilibrium, 
we get
\begin{equation}
\eta^{*}_{{\cancel{\scriptscriptstyle{\text{ TRS}}}}} = \gamma \eta_C + 
\frac{\gamma (1-\gamma)}{2} 
\eta_{C}^{2} + 
{\cal O} [\eta_{\rm C}^{3} ].
\label{seriesbtrs}
\end{equation}
The above series generalizes Eq. (\ref{etab}) which is obtained
with $x=1$, for which $\gamma  = \beta$.
For the case of optimal EMP where $\hat{\gamma} =4/7$ ($x=4/3$
\cite{Seifert2013}), the series
expansion (\ref{seriesbtrs}) is given by:
\begin{equation}
\eta^{*}_{\cancel{{\scriptscriptstyle{\text {TRS}}}}} =
\frac{4}{7}\eta_{\rm C}^{} + \frac{6}{49} 
\eta_{\rm C}^{2} + 
{\cal O} [\eta_{\rm C}^{3}].
\end{equation}
Thus, corresponding
to $\{ 1/2, 1/8 \} \equiv   \{ 4/8, 6/48 \}$  pair of 
universal coefficients for optimal EMP in the time-symmetric 
case, we obtain $\{4/7, 6/49 \}$ as the 
corresponding universal pair in the case of  broken TRS. 

\begin{figure}
 \includegraphics[width=8cm]{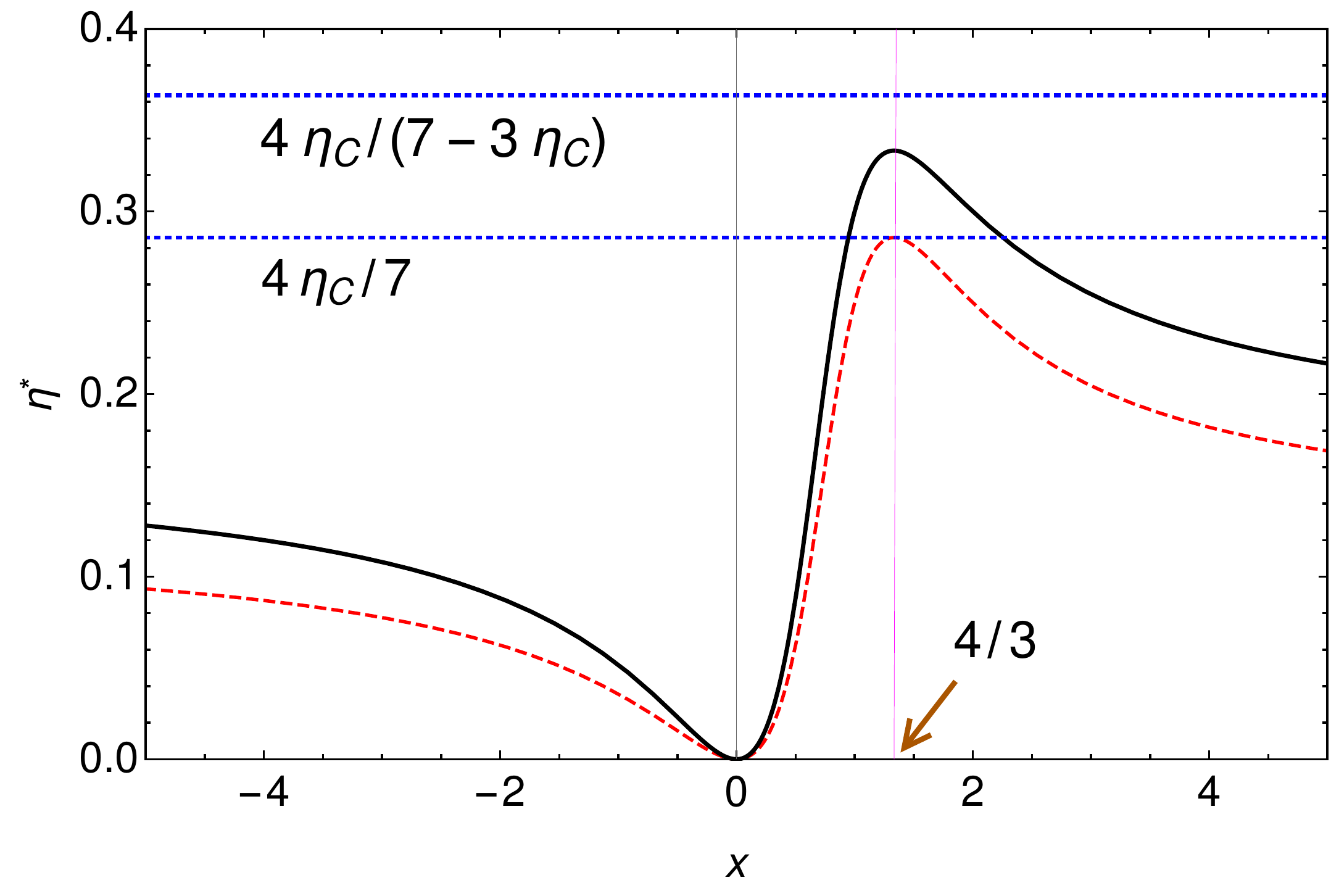}
 \caption{The (red) dashed curve denotes EMP for $T_0 = T_c$ whose
 optimal value is $4 \eta_{\rm C}^{}/7$, obtained at $x=4/3$ \cite{Seifert2013}.
 For $T_c \leq T_0 \leq T_h$, the optimal EMP is bounded between
 the two dashed horizontal lines, and is able to breach the $4 \eta_{\rm 
C}^{}/7$ value. 
 The thick (black) curve is the EMP for $T_0 = (T_h +T_c)/2$, which is 
also optimal at $x=4/3$. }
\end{figure}

{\it Model of coupled refrigerators}:
By reversing the energy flows in Fig. 1, we can study 
 two tightly coupled refrigerators in a similar manner. 
In this case, it is possible to optimize cooling power
of the total machine, as we show below.
\par\noindent
We can write the total rate of entropy generation as:
\begin{align}
\dot{S} 
&= \frac{\dot{W}}{T_0} - {\dot{Q}_{\rm av}}
\left(\frac{1}{T_c}-\frac{1}{T_h}\right).
\label{sdotavr}
\end{align}
Then, we identify the following flux-force pairs:
\begin{align}
{J}_1 &= \dot{x}, \qquad X_1 =  \frac{F}{T_0},
\label{j3macr}\\
{J}_2 &= \dot{Q}_{\rm av}, \quad X_2 = -\left( 
\frac{1}{T_c} - \frac{1}{T_h} \right ),
\label{j4macr}
\end{align}
so that $\dot{S}$ is cast in a bilinear form,  
$\dot{ S} \equiv {J}_1 X_1 + {J}_2 X_2$.

Within the linear-irreversible framework, the fluxes in Eqs. (\ref{j3macr})
and (\ref{j4macr}) take the 
following form:
\begin{align}
\dot{x}  &= {L}_{11} \frac{F}{T_0} + {L}_{12} X_2
\label{j1glor} \\
\dot{Q}_{\rm av} &= {L}_{12}\frac{F}{T_0}  + {L}_{22} X_2
\label{j2glor}
\end{align}
Then, we can derive the following relations:
\begin{align}
\dot{Q}_{h} &= {L}_{12}\frac{F}{T_0}  + {L}_{22} X_2
+ \omega \dot{W},
\label{qhr} \\
\dot{Q}_{c} &= {L}_{12}\frac{F}{T_0}  + {L}_{22} X_2
-(1-\omega) \dot{W}.
\label{qcr}
\end{align}
{\it Maximum cooling power}:
We maximize the cooling power, by setting 
\begin{equation}
\frac{\partial \dot{Q}_c}{ \partial F} = 0.
\label{qcmax}
\end{equation}
The optimal value of $F \equiv \hat{F}$ is given by:
\begin{equation}
\hat{F} = \frac{L_{12}X_2 T_0(2T_h-T_0)}{2 L_{11}(T_0-T_h)}.
\label{fhat}
\end{equation}
The coefficient of performance (COP) of the refrigerator
at maximum cooling power is defined as:
\begin{equation}
\xi^* = \frac{\dot{Q}_c (\hat{F})}{\dot{W}(\hat{F})},
\label{zist}
\end{equation}
and is evaluated to be:
\begin{equation}
\xi^* = \frac{(1-t_0)} {t_{0}^{2}(2-t_0)}
\left[ (2-t_0)^2 - \frac{4(1-t_0)}{q^2} \right],
\label{zimc}
\end{equation}
where $t_0 = T_0/T_h$. For a given $q$ value,
$\xi^*$ is a monotonic decreasing function of $t_0$.
So, the bounds of $\xi^*$ are given as:
\begin{equation}
0 \leq \xi^* \leq \frac{1}{\xi_{\rm C}(2+\xi_{\rm C})}  
\left[{(2+\xi_{\rm C})^2- 
\frac{4(1+\xi_{\rm C})}{q^2}}\right],
\end{equation}
where $ \xi_{\rm C} = T_c/(T_h - T_c)$ is the Carnot
bound for COP. For models with SC, Eq. (\ref{zimc})
gets simplified to: 
$\xi^* = \xi_{\rm C} (1-t_0)/(2-t_0)$, with the 
bounds as: $0 \leq \xi^* \leq \xi_{\rm C} /(2+\xi_{\rm C})$.

Concluding, we have studied the global performance of two tightly coupled 
engines within a three-reservoirs set up.
Assuming a linear-irreversible description  
where  the total rate of entropy generation
is defined in terms of a weighted average of the hot and cold
fluxes, we have optimized the total power and 
analysed the properties of the corresponding
efficiency at maximum power. The EMP, in general, 
depends on two ratios involving the three reservoir
temperatures. However, an interesting simplification 
occurs if the third temperature
is chosen as an algebraic mean between the hot
and cold temperatures. In this situation, the EMP can be
expressed in terms of the Carnot efficiency of 
the total set up, or equivalently, the ratio of cold to hot temperatures.
Further, the choice of this mean in the form of some common means
(such as geometric mean, harmonic mean and so on) yields 
well-known expressions for EMP found in previous studies
on two-reservoir set ups. Similarly, the universal
properties of EMP found in the latter case can also 
be identified in the three-reservoir scenario, when
the third temperature is a symmetric mean of hot and cold temperatures. 

Finally, universal features of EMP, surprising as they are,  
may be looked upon as a signature of the universality of  
thermodynamic approach. The present framework
for the global performance of coupled machines
provides an effective parameter in  $T_0$
which may be tuned to obtain EMP in a desired form,
thus bringing various mathematical forms of EMP under one formalism.
Such an approach, apart from providing a unified
viewpoint, can be instrumental in predicting novel features such as the 
6/49 second-order coefficient for EMP in the case of broken TRS. 
The generality of  thermodynamics deems 
it feasible that these features may 
be observed in systems with 
broken TRS, such as thermoelectric machines placed in 
an external magnetic field.
%%%%%%%%%%%%%%%%%%%%%%%%%%%%%%%%%%%%%%%%%%
%%%%%%%%%%%%%%%%%%%%%%%%%%%%%%%%%%%%%%%%%%

%
%

\begin{thebibliography}{30}%
\makeatletter
\providecommand \@ifxundefined [1]{%
 \@ifx{#1\undefined}
}%
\providecommand \@ifnum [1]{%
 \ifnum #1\expandafter \@firstoftwo
 \else \expandafter \@secondoftwo
 \fi
}%
\providecommand \@ifx [1]{%
 \ifx #1\expandafter \@firstoftwo
 \else \expandafter \@secondoftwo
 \fi
}%
\providecommand \natexlab [1]{#1}%
\providecommand \enquote  [1]{``#1''}%
\providecommand \bibnamefont  [1]{#1}%
\providecommand \bibfnamefont [1]{#1}%
\providecommand \citenamefont [1]{#1}%
\providecommand \href@noop [0]{\@secondoftwo}%
\providecommand \href [0]{\begingroup \@sanitize@url \@href}%
\providecommand \@href[1]{\@@startlink{#1}\@@href}%
\providecommand \@@href[1]{\endgroup#1\@@endlink}%
\providecommand \@sanitize@url [0]{\catcode `\\12\catcode `\$12\catcode
  `\&12\catcode `\#12\catcode `\^12\catcode `\_12\catcode `\%12\relax}%
\providecommand \@@startlink[1]{}%
\providecommand \@@endlink[0]{}%
\providecommand \url  [0]{\begingroup\@sanitize@url \@url }%
\providecommand \@url [1]{\endgroup\@href {#1}{\urlprefix }}%
\providecommand \urlprefix  [0]{URL }%
\providecommand \Eprint [0]{\href }%
\providecommand \doibase [0]{https://doi.org/}%
\providecommand \selectlanguage [0]{\@gobble}%
\providecommand \bibinfo  [0]{\@secondoftwo}%
\providecommand \bibfield  [0]{\@secondoftwo}%
\providecommand \translation [1]{[#1]}%
\providecommand \BibitemOpen [0]{}%
\providecommand \bibitemStop [0]{}%
\providecommand \bibitemNoStop [0]{.\EOS\space}%
\providecommand \EOS [0]{\spacefactor3000\relax}%
\providecommand \BibitemShut  [1]{\csname bibitem#1\endcsname}%
\let\auto@bib@innerbib\@empty
%</preamble>
\bibitem [{\citenamefont {Onsager}(1931)}]{Onsager1931a}%
  \BibitemOpen
  \bibfield  {author} {\bibinfo {author} {\bibfnamefont {L.}~\bibnamefont
  {Onsager}},\ }\bibfield  {title} {\bibinfo {title} {Reciprocal relations in
  irreversible processes. i.},\ }\href {https://doi.org/10.1103/PhysRev.37.405}
  {\bibfield  {journal} {\bibinfo  {journal} {Phys. Rev.}\ }\textbf {\bibinfo
  {volume} {37}},\ \bibinfo {pages} {405} (\bibinfo {year} {1931})}\BibitemShut
  {NoStop}%
\bibitem [{\citenamefont {Van~den Broeck}(2005)}]{Broeck2005}%
  \BibitemOpen
  \bibfield  {author} {\bibinfo {author} {\bibfnamefont {C.}~\bibnamefont
  {Van~den Broeck}},\ }\bibfield  {title} {\bibinfo {title} {Thermodynamic
  efficiency at maximum power},\ }\href@noop {} {\bibfield  {journal} {\bibinfo
   {journal} {Phys. Rev. Lett.}\ }\textbf {\bibinfo {volume} {95}},\ \bibinfo
  {pages} {190602} (\bibinfo {year} {2005})}\BibitemShut {NoStop}%
\bibitem [{\citenamefont {Curzon}\ and\ \citenamefont
  {Ahlborn}(1975)}]{Curzon1975}%
  \BibitemOpen
  \bibfield  {author} {\bibinfo {author} {\bibfnamefont {F.~L.}\ \bibnamefont
  {Curzon}}\ and\ \bibinfo {author} {\bibfnamefont {B.}~\bibnamefont
  {Ahlborn}},\ }\bibfield  {title} {\bibinfo {title} {Efficiency of a {C}arnot
  engine at maximum power output},\ }\href@noop {} {\bibfield  {journal}
  {\bibinfo  {journal} {Am. J. Phys.}\ }\textbf {\bibinfo {volume} {43}},\
  \bibinfo {pages} {22} (\bibinfo {year} {1975})}\BibitemShut {NoStop}%
\bibitem [{\citenamefont {Chen}\ and\ \citenamefont {Yan}(1989)}]{Chen1989}%
  \BibitemOpen
  \bibfield  {author} {\bibinfo {author} {\bibfnamefont {L.}~\bibnamefont
  {Chen}}\ and\ \bibinfo {author} {\bibfnamefont {Z.}~\bibnamefont {Yan}},\
  }\bibfield  {title} {\bibinfo {title} {The effect of heat‐transfer law on
  performance of a two‐heat‐source endoreversible cycle},\ }\href@noop {}
  {\bibfield  {journal} {\bibinfo  {journal} {J. Chem. Phys.}\ }\textbf
  {\bibinfo {volume} {90}},\ \bibinfo {pages} {3740} (\bibinfo {year}
  {1989})}\BibitemShut {NoStop}%
\bibitem [{\citenamefont {Schmiedl}\ and\ \citenamefont
  {Seifert}(2008)}]{Schmiedl2008}%
  \BibitemOpen
  \bibfield  {author} {\bibinfo {author} {\bibfnamefont {T.}~\bibnamefont
  {Schmiedl}}\ and\ \bibinfo {author} {\bibfnamefont {U.}~\bibnamefont
  {Seifert}},\ }\bibfield  {title} {\bibinfo {title} {Efficiency at maximum
  power: An analytically solvable model for stochastic heat engines},\
  }\href@noop {} {\bibfield  {journal} {\bibinfo  {journal} {EPL (Europhysics
  Letters)}\ }\textbf {\bibinfo {volume} {81}},\ \bibinfo {pages} {20003}
  (\bibinfo {year} {2008})}\BibitemShut {NoStop}%
\bibitem [{\citenamefont {Esposito}\ \emph
  {et~al.}(2010{\natexlab{a}})\citenamefont {Esposito}, \citenamefont {Kawai},
  \citenamefont {Lindenberg},\ and\ \citenamefont {Van~den
  Broeck}}]{Esposito2010prl}%
  \BibitemOpen
  \bibfield  {author} {\bibinfo {author} {\bibfnamefont {M.}~\bibnamefont
  {Esposito}}, \bibinfo {author} {\bibfnamefont {R.}~\bibnamefont {Kawai}},
  \bibinfo {author} {\bibfnamefont {K.}~\bibnamefont {Lindenberg}},\ and\
  \bibinfo {author} {\bibfnamefont {C.}~\bibnamefont {Van~den Broeck}},\
  }\bibfield  {title} {\bibinfo {title} {Efficiency at maximum power of
  low-dissipation carnot engines},\ }\href@noop {} {\bibfield  {journal}
  {\bibinfo  {journal} {Phys. Rev. Lett.}\ }\textbf {\bibinfo {volume} {105}},\
  \bibinfo {pages} {150603} (\bibinfo {year} {2010}{\natexlab{a}})}\BibitemShut
  {NoStop}%
\bibitem [{\citenamefont {Moreau}\ \emph {et~al.}(2012)\citenamefont {Moreau},
  \citenamefont {Gaveau},\ and\ \citenamefont {Schulman}}]{Moreau2012}%
  \BibitemOpen
  \bibfield  {author} {\bibinfo {author} {\bibfnamefont {M.}~\bibnamefont
  {Moreau}}, \bibinfo {author} {\bibfnamefont {B.}~\bibnamefont {Gaveau}},\
  and\ \bibinfo {author} {\bibfnamefont {L.~S.}\ \bibnamefont {Schulman}},\
  }\bibfield  {title} {\bibinfo {title} {Efficiency of a thermodynamic motor at
  maximum power},\ }\href {https://doi.org/10.1103/PhysRevE.85.021129}
  {\bibfield  {journal} {\bibinfo  {journal} {Phys. Rev. E}\ }\textbf {\bibinfo
  {volume} {85}},\ \bibinfo {pages} {021129} (\bibinfo {year}
  {2012})}\BibitemShut {NoStop}%
\bibitem [{\citenamefont {Benenti}\ \emph {et~al.}(2011)\citenamefont
  {Benenti}, \citenamefont {Saito},\ and\ \citenamefont
  {Casati}}]{Benenti2011}%
  \BibitemOpen
  \bibfield  {author} {\bibinfo {author} {\bibfnamefont {G.}~\bibnamefont
  {Benenti}}, \bibinfo {author} {\bibfnamefont {K.}~\bibnamefont {Saito}},\
  and\ \bibinfo {author} {\bibfnamefont {G.}~\bibnamefont {Casati}},\
  }\bibfield  {title} {\bibinfo {title} {Thermodynamic bounds on efficiency for
  systems with broken time-reversal symmetry},\ }\href
  {https://doi.org/10.1103/PhysRevLett.106.230602} {\bibfield  {journal}
  {\bibinfo  {journal} {Phys. Rev. Lett.}\ }\textbf {\bibinfo {volume} {106}},\
  \bibinfo {pages} {230602} (\bibinfo {year} {2011})}\BibitemShut {NoStop}%
\bibitem [{\citenamefont {Brandner}\ \emph {et~al.}(2013)\citenamefont
  {Brandner}, \citenamefont {Saito},\ and\ \citenamefont
  {Seifert}}]{Seifert2013}%
  \BibitemOpen
  \bibfield  {author} {\bibinfo {author} {\bibfnamefont {K.}~\bibnamefont
  {Brandner}}, \bibinfo {author} {\bibfnamefont {K.}~\bibnamefont {Saito}},\
  and\ \bibinfo {author} {\bibfnamefont {U.}~\bibnamefont {Seifert}},\
  }\bibfield  {title} {\bibinfo {title} {Strong bounds on onsager coefficients
  and efficiency for three-terminal thermoelectric transport in a magnetic
  field},\ }\href {https://doi.org/10.1103/PhysRevLett.110.070603} {\bibfield
  {journal} {\bibinfo  {journal} {Phys. Rev. Lett.}\ }\textbf {\bibinfo
  {volume} {110}},\ \bibinfo {pages} {070603} (\bibinfo {year}
  {2013})}\BibitemShut {NoStop}%
\bibitem [{\citenamefont {Izumida}\ and\ \citenamefont
  {Okuda}(2012)}]{Izumida2012}%
  \BibitemOpen
  \bibfield  {author} {\bibinfo {author} {\bibfnamefont {Y.}~\bibnamefont
  {Izumida}}\ and\ \bibinfo {author} {\bibfnamefont {K.}~\bibnamefont
  {Okuda}},\ }\bibfield  {title} {\bibinfo {title} {Efficiency at maximum power
  of minimally nonlinear irreversible heat engines},\ }\href
  {https://doi.org/10.1209/0295-5075/97/10004} {\bibfield  {journal} {\bibinfo
  {journal} {{EPL} (Europhysics Letters)}\ }\textbf {\bibinfo {volume} {97}},\
  \bibinfo {pages} {10004} (\bibinfo {year} {2012})}\BibitemShut {NoStop}%
\bibitem [{\citenamefont {den Broeck}(2013)}]{Broeck2013}%
  \BibitemOpen
  \bibfield  {author} {\bibinfo {author} {\bibfnamefont {C.~V.}\ \bibnamefont
  {den Broeck}},\ }\bibfield  {title} {\bibinfo {title} {Efficiency at maximum
  power in the low-dissipation limit},\ }\href
  {https://doi.org/10.1209/0295-5075/101/10006} {\bibfield  {journal} {\bibinfo
   {journal} {{EPL} (Europhysics Letters)}\ }\textbf {\bibinfo {volume}
  {101}},\ \bibinfo {pages} {10006} (\bibinfo {year} {2013})}\BibitemShut
  {NoStop}%
\bibitem [{\citenamefont {Johal}(2018)}]{Johal2018}%
  \BibitemOpen
  \bibfield  {author} {\bibinfo {author} {\bibfnamefont {R.~S.}\ \bibnamefont
  {Johal}},\ }\bibfield  {title} {\bibinfo {title} {Global linear-irreversible
  principle for optimization in finite-time thermodynamics},\ }\href
  {http://stacks.iop.org/0295-5075/121/i=5/a=50009} {\bibfield  {journal}
  {\bibinfo  {journal} {EPL (Europhysics Letters)}\ }\textbf {\bibinfo {volume}
  {121}},\ \bibinfo {pages} {50009} (\bibinfo {year} {2018})}\BibitemShut
  {NoStop}%
\bibitem [{\citenamefont {Esposito}\ \emph {et~al.}(2009)\citenamefont
  {Esposito}, \citenamefont {Lindenberg},\ and\ \citenamefont {Van~den
  Broeck}}]{Lindenberg2009}%
  \BibitemOpen
  \bibfield  {author} {\bibinfo {author} {\bibfnamefont {M.}~\bibnamefont
  {Esposito}}, \bibinfo {author} {\bibfnamefont {K.}~\bibnamefont
  {Lindenberg}},\ and\ \bibinfo {author} {\bibfnamefont {C.}~\bibnamefont
  {Van~den Broeck}},\ }\bibfield  {title} {\bibinfo {title} {Universality of
  efficiency at maximum power},\ }\href@noop {} {\bibfield  {journal} {\bibinfo
   {journal} {Phys. Rev. Lett.}\ }\textbf {\bibinfo {volume} {102}},\ \bibinfo
  {pages} {130602} (\bibinfo {year} {2009})}\BibitemShut {NoStop}%
\bibitem [{\citenamefont {Esposito}\ \emph
  {et~al.}(2010{\natexlab{b}})\citenamefont {Esposito}, \citenamefont {Kawai},
  \citenamefont {Lindenberg},\ and\ \citenamefont {Van~den
  Broeck}}]{Lindenberg2010}%
  \BibitemOpen
  \bibfield  {author} {\bibinfo {author} {\bibfnamefont {M.}~\bibnamefont
  {Esposito}}, \bibinfo {author} {\bibfnamefont {R.}~\bibnamefont {Kawai}},
  \bibinfo {author} {\bibfnamefont {K.}~\bibnamefont {Lindenberg}},\ and\
  \bibinfo {author} {\bibfnamefont {C.}~\bibnamefont {Van~den Broeck}},\
  }\bibfield  {title} {\bibinfo {title} {Quantum-dot {C}arnot engine at maximum
  power},\ }\href@noop {} {\bibfield  {journal} {\bibinfo  {journal} {Phys.
  Rev. E}\ }\textbf {\bibinfo {volume} {81}} (\bibinfo {year}
  {2010}{\natexlab{b}})}\BibitemShut {NoStop}%
\bibitem [{\citenamefont {Johal}(2019)}]{Johal2019}%
  \BibitemOpen
  \bibfield  {author} {\bibinfo {author} {\bibfnamefont {R.~S.}\ \bibnamefont
  {Johal}},\ }\bibfield  {title} {\bibinfo {title} {Performance optimization of
  low-dissipation thermal machines revisited},\ }\href
  {https://doi.org/10.1103/PhysRevE.100.052101} {\bibfield  {journal} {\bibinfo
   {journal} {Phys. Rev. E}\ }\textbf {\bibinfo {volume} {100}},\ \bibinfo
  {pages} {052101} (\bibinfo {year} {2019})}\BibitemShut {NoStop}%
  \bibitem{comment_Johal2018} An effective thermodynamic approach 
was studied for finite-time discrete heat engines 
by one of the authors in Ref. \cite{Johal2018}.  
  \bibitem{comment_car}
As the two engines are tightly coupled to each other, so the 
efficiencies of the first and the second engines are related to
the global efficiency as, $\eta = 1- (1-\eta_1)(1-\eta_2)$.
Thus, $\eta_{\rm C}^{}$ is still a measure for the maximal
efficiency of the global system, which is achieved when both 
engines run reversibly.
\bibitem [{\citenamefont {Johal}\ and\ \citenamefont
  {Rai}(2016)}]{JohalRai2016}%
  \BibitemOpen
  \bibfield  {author} {\bibinfo {author} {\bibfnamefont {R.~S.}\ \bibnamefont
  {Johal}}\ and\ \bibinfo {author} {\bibfnamefont {R.}~\bibnamefont {Rai}},\
  }\bibfield  {title} {\bibinfo {title} {Near-equilibrium universality and
  bounds on efficiency in quasi-static regime with finite source and sink},\
  }\href@noop {} {\bibfield  {journal} {\bibinfo  {journal} {EPL (Europhysics
  Letters)}\ }\textbf {\bibinfo {volume} {113}},\ \bibinfo {pages} {10006}
  (\bibinfo {year} {2016})}\BibitemShut {NoStop}%
\bibitem [{\citenamefont {Singh}\ and\ \citenamefont
  {Johal}(2018)}]{VarinderJohal2018}%
  \BibitemOpen
  \bibfield  {author} {\bibinfo {author} {\bibfnamefont {V.}~\bibnamefont
  {Singh}}\ and\ \bibinfo {author} {\bibfnamefont {R.~S.}\ \bibnamefont
  {Johal}},\ }\bibfield  {title} {\bibinfo {title} {Feynman–{S}moluchowski
  engine at high temperatures and the role of constraints},\ }\href
  {http://stacks.iop.org/1742-5468/2018/i=7/a=073205} {\bibfield  {journal}
  {\bibinfo  {journal} {J. Stat. Mech.}\ }\textbf {\bibinfo {volume} {2018}},\
  \bibinfo {pages} {073205} (\bibinfo {year} {2018})}\BibitemShut {NoStop}%
\bibitem [{\citenamefont {Tu}(2008)}]{Tu2008}%
  \BibitemOpen
  \bibfield  {author} {\bibinfo {author} {\bibfnamefont {Z.~C.}\ \bibnamefont
  {Tu}},\ }\bibfield  {title} {\bibinfo {title} {Efficiency at maximum power of
  {F}eynman's ratchet as a heat engine},\ }\href@noop {} {\bibfield  {journal}
  {\bibinfo  {journal} {J. Phys. A: Math. and Theor.}\ }\textbf {\bibinfo
  {volume} {41}},\ \bibinfo {pages} {312003} (\bibinfo {year}
  {2008})}\BibitemShut {NoStop}%
\bibitem [{\citenamefont {Van~den Broeck}\ and\ \citenamefont
  {Lindenberg}(2012)}]{Broeck2012}%
  \BibitemOpen
  \bibfield  {author} {\bibinfo {author} {\bibfnamefont {C.}~\bibnamefont
  {Van~den Broeck}}\ and\ \bibinfo {author} {\bibfnamefont {K.}~\bibnamefont
  {Lindenberg}},\ }\bibfield  {title} {\bibinfo {title} {Efficiency at maximum
  power for classical particle transport},\ }\href@noop {} {\bibfield
  {journal} {\bibinfo  {journal} {Phys. Rev. E}\ }\textbf {\bibinfo {volume}
  {86}},\ \bibinfo {pages} {041144} (\bibinfo {year} {2012})}\BibitemShut
  {NoStop}%
\bibitem [{\citenamefont {Wang}\ \emph {et~al.}(2013)\citenamefont {Wang},
  \citenamefont {Wang}, \citenamefont {He},\ and\ \citenamefont
  {Ma}}]{WangTLS2013}%
  \BibitemOpen
  \bibfield  {author} {\bibinfo {author} {\bibfnamefont {R.}~\bibnamefont
  {Wang}}, \bibinfo {author} {\bibfnamefont {J.}~\bibnamefont {Wang}}, \bibinfo
  {author} {\bibfnamefont {J.}~\bibnamefont {He}},\ and\ \bibinfo {author}
  {\bibfnamefont {Y.}~\bibnamefont {Ma}},\ }\bibfield  {title} {\bibinfo
  {title} {Efficiency at maximum power of a heat engine working with a
  two-level atomic system},\ }\href@noop {} {\bibfield  {journal} {\bibinfo
  {journal} {Phys. Rev. E}\ }\textbf {\bibinfo {volume} {87}},\ \bibinfo
  {pages} {042119} (\bibinfo {year} {2013})}\BibitemShut {NoStop}%
\bibitem [{\citenamefont {Erdman}\ \emph {et~al.}(2017)\citenamefont {Erdman},
  \citenamefont {Mazza}, \citenamefont {Bosisio}, \citenamefont {Benenti},
  \citenamefont {Fazio},\ and\ \citenamefont {Taddei}}]{Erdman2017}%
  \BibitemOpen
  \bibfield  {author} {\bibinfo {author} {\bibfnamefont {P.~A.}\ \bibnamefont
  {Erdman}}, \bibinfo {author} {\bibfnamefont {F.}~\bibnamefont {Mazza}},
  \bibinfo {author} {\bibfnamefont {R.}~\bibnamefont {Bosisio}}, \bibinfo
  {author} {\bibfnamefont {G.}~\bibnamefont {Benenti}}, \bibinfo {author}
  {\bibfnamefont {R.}~\bibnamefont {Fazio}},\ and\ \bibinfo {author}
  {\bibfnamefont {F.}~\bibnamefont {Taddei}},\ }\bibfield  {title} {\bibinfo
  {title} {Thermoelectric properties of an interacting quantum dot based heat
  engine},\ }\href {https://doi.org/10.1103/PhysRevB.95.245432} {\bibfield
  {journal} {\bibinfo  {journal} {Phys. Rev. B}\ }\textbf {\bibinfo {volume}
  {95}},\ \bibinfo {pages} {245432} (\bibinfo {year} {2017})}\BibitemShut
  {NoStop}%
\bibitem [{\citenamefont {Cavina}\ \emph {et~al.}(2017)\citenamefont {Cavina},
  \citenamefont {Mari},\ and\ \citenamefont {Giovannetti}}]{Cavina2017}%
  \BibitemOpen
  \bibfield  {author} {\bibinfo {author} {\bibfnamefont {V.}~\bibnamefont
  {Cavina}}, \bibinfo {author} {\bibfnamefont {A.}~\bibnamefont {Mari}},\ and\
  \bibinfo {author} {\bibfnamefont {V.}~\bibnamefont {Giovannetti}},\
  }\bibfield  {title} {\bibinfo {title} {Slow dynamics and thermodynamics of
  open quantum systems},\ }\href
  {https://doi.org/10.1103/PhysRevLett.119.050601} {\bibfield  {journal}
  {\bibinfo  {journal} {Phys. Rev. Lett.}\ }\textbf {\bibinfo {volume} {119}},\
  \bibinfo {pages} {050601} (\bibinfo {year} {2017})}\BibitemShut {NoStop}%
\bibitem{comment_taylor}
For example, if $T_0 = [(1-p_1)/T_h + p_1/T_c]^{-1}$, 
a weighted harmonic mean ($0\leq p_1 \leq 1$), then 
${\cal M}(1,1-\eta_{\rm C}) = 1-p_1 \eta_{\rm C} 
+ p_1(p_1-1) \eta_{\rm C}^{2} + O[\eta_{\rm C}^{3}]$,
so that $a_1 = p_1$, $a_2 = p_1 (p_1 -1)$, and so on.  
\bibitem{comment_a1}
Many well-known means are symmetric means, such as
arithmetic mean  $(a+b)/2$, geometric mean $\sqrt{a b}$,
and harmonic mean $2 a b/(a+b)$. Further, these means 
satisfy the following property:
 \begin{equation}
\left. \frac{\partial {\cal M} }{\partial a}\right \vert_{a=b}  = 
\left. \frac{\partial {\cal M}}{\partial b}\right \vert_{b=a} = \frac{1}{2},
\nonumber
\end{equation}
which implies $a_1 = 1/2$. 
\bibitem [{\citenamefont {Stolarsky}(1975)}]{Stolarsky1975}%
  \BibitemOpen
  \bibfield  {author} {\bibinfo {author} {\bibfnamefont {K.~B.}\ \bibnamefont
  {Stolarsky}},\ }\bibfield  {title} {\bibinfo {title} {Generalizations of the
  logarithmic mean},\ }\href {http://www.jstor.org/stable/2689825} {\bibfield
  {journal} {\bibinfo  {journal} {Mathematics Magazine}\ }\textbf {\bibinfo
  {volume} {48}},\ \bibinfo {pages} {87} (\bibinfo {year} {1975})}\BibitemShut
  {NoStop}%
\bibitem [{\citenamefont {Leach}\ and\ \citenamefont
  {Sholander}(1978)}]{Leach1978}%
  \BibitemOpen
  \bibfield  {author} {\bibinfo {author} {\bibfnamefont {E.~B.}\ \bibnamefont
  {Leach}}\ and\ \bibinfo {author} {\bibfnamefont {M.~C.}\ \bibnamefont
  {Sholander}},\ }\bibfield  {title} {\bibinfo {title} {Extended mean values},\
  }\href {http://www.jstor.org/stable/2321783} {\bibfield  {journal} {\bibinfo
  {journal} {The American Mathematical Monthly}\ }\textbf {\bibinfo {volume}
  {85}},\ \bibinfo {pages} {84} (\bibinfo {year} {1978})}\BibitemShut {NoStop}%
  \end{thebibliography}
\end{document}